\documentclass[preprint,prb,draft,showkeys,showpacs,tightenlines]{revtex4}
\begin{document}
\preprint{Bicocca-FT-02-14  July 2002}

\title
 {Critical parameters  and universal amplitude ratios\\
  of two-dimensional spin-$S$ Ising models \\
  using high- and low-temperature expansions\\}

\author{P. Butera\cite{pb} and M. Comi}
\affiliation
{Istituto Nazionale di Fisica Nucleare and\\
Dipartimento di Fisica, Universit\`a di Milano-Bicocca\\
 3 Piazza della Scienza, 20126 Milano, Italy}
\author{A. J. Guttmann\cite{tg}}
\affiliation
{Department of Mathematics and Statistics, The University
of Melbourne, Victoria, 3010, Australia.}
\date{\today}
\begin{abstract}

For the study of Ising models  of general spin $S$  on the square 
lattice, we have  combined our recently extended 
 high-temperature expansions   with the low-temperature
 expansions derived some time ago by Enting, Guttmann and Jensen. 
We have computed for the first time various  
critical parameters and improved the estimates of others. 
 Moreover the properties of hyperscaling and of
 universality  (spin $S$ independence) of exponents and 
of various dimensionless amplitude combinations 
have been verified accurately.
Assuming the validity of the lattice-lattice scaling, 
from our estimates of critical amplitudes for the square lattice 
 we have also obtained estimates of the corresponding amplitudes 
for the spin $S$ Ising model
on the triangular, honeycomb, and kagom\'e lattices.
\end{abstract}


\maketitle

\section{Introduction}
The properties of the
spin $S=1/2$ two-dimensional Ising model  with nearest-neighbor 
interactions  in zero magnetic field, 
 have been extensively explored in the last six decades.
Much more modest efforts have been devoted to the 
study of the simplest generalizations of the model 
to spin $S>1/2$. The main reason is probably that 
 these models are not  known to be  solvable  
 or, at least, to have any simple duality 
 property which can help to extend 
the  small body of information coming from
numerical  methods of limited accuracy such as stochastic simulations,
series expansions or  transfer-matrix calculations.

The first important result from a comparative study of Ising models 
 for different values of the spin came from
  pioneering work by  Domb and Sykes \cite{dosy}.
  They analysed the high-temperature(HT) expansion of
 the susceptibility $\chi(K;S)$ through $O(K^6)$ 
in the three-dimensional case and conjectured
  that the value of the critical exponent $\gamma(S)$ is
 independent of the  spin magnitude. 
This was the first step towards the modern formulation of the 
critical-universality hypothesis.
 Similar analyses were soon  repeated by other authors 
  using both HT\cite{moore,camp}
and  low-temperature (LT) expansions\cite{foxgu} for  two-dimensional
 systems.
 Unfortunately,  the series derived in those years 
 were  rather short and therefore the results
 of the analyses could not reach a sufficient 
accuracy or were inconclusive.
 It was only in  1980 that  Nickel\cite{nick21,nr90}
 finally  
extended through $O(K^{21})$ the HT series in two dimensions 
on  the square (sq) lattice and in three dimensions 
on the body-centered cubic lattice. 
 The expansions of $\chi(K;S)$ and of the second moment of 
the spin-spin correlation function $\mu_2(K;S)$ 
 were then published only for $S=1/2,1,2,\infty$.
More recently also the LT 
expansions on the sq lattice for $S=1/2,1,3/2,2,5/2,3$ 
were considerably extended  
 by Enting, Jensen and Guttmann\cite{ejg}.
We have summarized in Table \ref{tab1} and  Table \ref{tab2} the state
 of HT\cite{camp,bakin} and LT\cite{ejg}   expansions before our work.

\begin{table}[b]
\squeezetable
\caption{ The longest HT   expansions, 
published (or obtainable from data in the literature)
 before our work\cite{bcesse25,bcol}, 
for the susceptibility $\chi(K;S)$,  the second moment of the 
correlation function $\mu_2(K;S)$  and  
the second field-derivative of the susceptiblity $\chi_4(K;S)$
  in the case of  the  Ising  models with general spin $S$ 
on  two-dimensional lattices. 
It should be noted that in the special case 
 $S=1/2$, on the sq lattice,
 much longer expansions\cite{orri} for $\chi$ and $\mu_2$
  have been computed. However, 
 the published expansions\cite{nick21}
  of $\chi_4(K;1/2)$ on the sq lattice  
do not extend beyond $K^{17}$. }
\label{tab1}
\begin{ruledtabular}
\begin{tabular}{lccc}
  Observable &  Lattice  & Order &  Ref. \\
\tableline

$\chi(K;S)$, $\mu_2(K;S)$ &sq &21& \cite{nr90}  \\

$\chi_4(K;S)$ &sq & 10& \cite{bakin} \\

$\chi(K;S)$, $\mu_2(K;S)$ &tr &10& \cite{camp}  \\

$\chi_4(K;S)$ &tr & 10& \cite{bakin} \\

\end{tabular}
\end{ruledtabular}
\end{table}

\begin{table}
\squeezetable
\caption{ The longest  LT expansions, 
presently available for the spontaneous magnetization,
 the specific heat 
and  the susceptibility 
  in the case of  the  Ising  models with various values of the 
 spin on the sq lattice. }
\label{tab2}
\begin{ruledtabular}
\begin{tabular}{lccc}
  Observable &  Lattice  & Order &  Ref. \\
\tableline
$M(u;1)$, $C(u;1)$, $\chi(u;1)$ &sq &113&\cite{ejg}\\
$M(u;3/2)$, $C(u;3/2)$, $\chi(u;3/2)$ &sq&100&\cite{ejg}\\
$M(u;2)$, $C(u;2)$, $\chi(u;2)$ &sq &119&\cite{ejg}\\
$M(u;5/2)$, $C(u;5/2)$, $\chi(u;5/2)$&sq&126&\cite{ejg}\\
$M(u;3)$, $C(u;3)$, $\chi(u;3)$ &sq &154& \cite{ejg}\\
\end{tabular}
\end{ruledtabular}
\end{table}

In spite of the very large number of LT expansion coefficients 
now available, the analysis of the  series remains arduous 
 due to occurrence\cite{ejg} of  numerous unphysical singularities 
in the complex temperature plane\cite{fisherzeroes} which are 
 closer to the origin than 
the physical singularity and whose
structure becomes increasingly complicated with $S$.
As a consequence, the LT study of Ref.\cite{ejg} has been   an alarming 
 lesson on the subtleties in the analysis
 of  slowly convergent series more than a source of 
 accurate estimates of the critical parameters of the  
 models. 

Many intriguing indications and  conjectures about the structure 
of these unphysical singularities  also came in the same period 
from work by Matveev and Shrock\cite{ms} 
who examined the spin $S$ models 
 on various two-dimensional lattices using transfer-matrix methods.
 
Here we discuss some  results of an analysis of  
HT series for the sq lattice 
recently extended\cite{bcesse25} 
by linked-cluster expansion  techniques. 

 For the nearest-neighbor correlation function $G(K;S)$, 
 for $\chi(K;S)$, and $\mu_2(K;S)$ our
 series reach order $K^{25}$,
  while  for the second field derivative of the 
susceptibility $\chi_4(K;S)$ they extend through  $O(K^{23})$. 
In order to make alternative analyses possible, 
our vast collection of series data both for two- and 
three-dimensional lattices was made easily accessible\cite{bcol} 
on-line for $S=1/2,1,3/2,2,5/2,3,7/2,4,5,\infty$. 
 It should be noted that HT and LT 
expansions as extensive as those obtained by 
Nickel et al. in Ref.\cite{nick21} and more recently 
in Ref.\cite{orri} (only for $\chi$) in the very special 
case of the (partially solvable) two-dimensional
 $S=1/2$ Ising model   seem presently beyond  reach  for  $S>1/2$.

The HT series show somewhat simpler and faster convergence 
properties than the LT series, because the  behavior 
of the coefficients is
 dominated by the physical singularity. 
 Although, even in this case,
 these favorable properties slightly deteriorate for  $S>2$, 
we can hope 
  to determine   basic HT
critical parameters with a reasonable accuracy  
for various values of $S$.
Moreover it is also worthwhile to reconsider the LT expansions 
of Ref.\cite{ejg} for the sq lattice,
 because by relying on the results of our HT analysis, we can
  improve some estimates of the LT critical parameters and thus obtain
 new determinations of  universal combinations
\cite{aha} of LT and HT amplitudes.
 No theoretical surprises are expected from this analysis,  
however we believe it is still useful
 to  improve the rather modest numerical 
precision presently available   
 even for basic critical parameters
  like the critical temperatures, to determine various 
critical amplitudes for which no estimates are yet known 
and to use our results  to test with higher
 accuracy the validity of hyperscaling
 and of  universality with respect to the magnitude of the spin. 

Almost all the computational effort in extending series for the
two-dimensional Ising model for $S > 1/2$ has been devoted to
square-lattice series. However by making use of the theory of
{\em lattice-lattice scaling}, as developed by Betts {\em et
al.}\cite{bgj}
and extended by Gaunt and Guttmann\cite{gg}, using our 
estimates of the critical amplitudes on the square lattice, 
we are able to calculate
the corresponding amplitudes on other two-dimensional lattices to
precisely the same precision as they are known for the square lattice.

\section{ The spin-$S$ Ising models}

  The spin-$S$ Ising models  with nearest-neighbor 
interaction are defined by  the Hamiltonian:

\begin{equation}
H \{ s \} = -{\frac {J} {2}} \sum_{( \vec x,{\vec x}') } 
s({\vec x})  s({\vec x}') -h\sum_{\vec x} s({\vec x})
\label{hamilt} \end{equation}

where $J$ is the exchange coupling, and
  $s(\vec x)=s^z(\vec x)/S$  with  $s^z(\vec x)$  a  
classical spin  variable at the
lattice site $\vec x$, taking the $2S+1$ values 
$-S, -S+1, \ldots,S-1, S$.    
 The sum runs over  all nearest-neighbor pairs of  sites. 
We shall restrict ourselves to the  square lattice and
 consider expansions either 
in the usual HT variable $K=J/k_BT$   and in the  natural LT variable 
  $u(S)=exp(-K/S^2)$. Here $T$ is the temperature, $k_B$
 the Boltzmann constant, and $K$ will be called
``inverse temperature'' for brevity. 
In the critical region we shall also refer to 
 the standard reduced-temperature variable 
 $t(S)=T/T_c(S)- 1 = K_c(S)/K - 1$.

 In the HT phase, the basic observables are the 
connected $2n$-spin correlation functions. Our series\cite{bcol} cover 
 quantities related to the  
 two-spin correlation functions 
$\langle s(\vec x)  s(\vec y) \rangle_c$
 and to the four-spin correlation functions 
$\langle  s(\vec x)  s(\vec y) s(\vec z)  s(\vec t)\rangle_{c}$.
 
In the LT phase the symmetry is broken and  the 
$n$-spin correlations are non trivial also for odd $n$.
In particular, we shall reconsider the LT expansions of  
 the magnetization, 
the susceptibility and the specific heat derived 
for $S=1,3/2,2,5/2,3$ in Ref.\cite{ejg}.

The spontaneous magnetization is defined by 

\begin{equation}
M(T;S)= \lim_{h \to  0+}  \langle s(\vec 0)\rangle.
\end{equation}

The internal energy per spin 
is given in terms of the nearest-neighbor correlation function by

\begin{equation}
U(T;S)=- \frac{qJ} {2}\langle s(\vec 0)  s(\vec \delta) \rangle
\end{equation}
where  $\vec \delta$ is a nearest-neighbor lattice vector and 
$q$ is the lattice coordination number.

The specific heat is  the temperature-derivative of the internal 
energy at fixed zero external field

\begin{equation}
C_H(T;S) =\frac {dU(T;S)} {dT}. 
\end{equation}

In terms of $\chi(T;S)$, the zero-field reduced susceptibility,

\begin{equation}
\chi(T;S) = \sum_{\vec x} \langle s(\vec 0)  s(\vec x) \rangle_c  
\label{chi} \end{equation}

and  of  $ \mu_{2}(T;S)$, the second  moment 
of the correlation function,

\begin{equation}
 \mu_{2}(T;S)=\sum_{\vec x} \vec x^2 \langle s(\vec 0)  s(\vec x) 
\rangle_c 
\end{equation}

 the  ``second-moment correlation length'' $\xi(T;S)$ is defined  by 

\begin{equation}
 \xi^{2}(T;S)= \frac  {\mu_{2}(T;S)} {4\chi(T;S) }.
\end{equation}

The second field-derivative of the
susceptibility $\chi_{4}(T;S)$ is defined by

\begin{equation}
 \chi_{4}(T;S)=  \sum_{\vec x,\vec y,\vec z}
\langle  s(\vec 0)  s(\vec x) s(\vec y)  s(\vec z)\rangle_{c}.
\label{chi4}\end{equation}

\section{ Definitions of critical parameters}

 In terms of the asymptotic behavior 
 of these observables, we can
 now define the critical parameters, amplitudes and exponents
 that we are going to estimate using  HT and LT series.

The spontaneous magnetization has the asymptotic behavior
\begin{equation}
M^{(-)}(T;S) \simeq B^{(-)}(S)|t(S)|^{\beta(S)}\Big(1+ 
a^{(-)}_M(S)|t(S)|^{\theta(S)} +..\Big)
\label{mag}\end{equation}

 as $t(S)  \rightarrow 0-$.

   The asymptotic 
behaviour of the susceptibility
 as $t(S) \rightarrow 0 \pm$, 
 is expected to be

\begin{equation}
 \chi^{(\pm)}(T;S)
\simeq C^{(\pm)}(S)|t(S)|^{-\gamma(S)}\Big(1+ a^{(\pm)}_{\chi}(S)
|t(S)|^{\theta(S)} + \ldots
+ b^{(\pm)}_{\chi}(S)t(S) + \ldots \Big)
\label{conf}\end{equation}

 The 
correlation length 

\begin{equation}
 \xi^{(\pm)}(T;S)
\simeq f^{(\pm)}(S)|t(S)|^{-\nu(S)}\Big(1
+ a^{(\pm)}_{\xi}(S)
|t(S)|^{\theta(S)} + \ldots
+ b^{(\pm)}_{\xi}(S)t(S) +\ldots \Big)
\label{confxi}\end{equation} 

 the specific heat

\begin{equation}
 C_H^{(\pm)}(T;S)/k_B
\simeq  A^{(\pm)}(S)\ln |t(S)|
\Big(1+ a_C^{(\pm)}(S)
|t(S)|^{\theta(S)} +\ldots
+ b_C^{(\pm)}(S)t(S) + \ldots \Big)
\label{confc}\end{equation}

and the second field-derivative of the 
susceptibility $\chi_{4}(K;S)$

\begin{equation}
 \chi^{(\pm)}_{4}(T;S)
\simeq -C^{(\pm)}_{4}(S)|t(S)|^{-\gamma_4(S)}\Big(1+ a^{(\pm)}_{4}(S)
|t(S)|^{\theta(S)} +\ldots
+ b^{(\pm)}_{4}(S)t(S) + \ldots \Big)
\label{conf4}\end{equation}  
have analogous asymptotic behaviours.

  Different (universal) critical
 exponents $\beta(S), \gamma(S), \nu(S), \gamma_4(S) $ 
and different (non-universal) critical amplitudes $B^{(-)}(S)$, 
 $C^{(\pm)}(S)$, $f^{(\pm)}(S) \ldots $, $a^{(\pm)}_{\chi}(S)$,
  $a^{(\pm)}_{\xi}(S)$, etc.  are associated with the 
various observables. We have reported in such  detail 
our definitions of the  critical amplitudes,
 because they differ significantly from those  
of other authors and it is 
 necessary to use  a consistent normalisation convention when comparing
 models expected to belong to the same universality class.
Let us notice in particular that 
 the estimates reported in the tables of Ref.\cite{ejg} for the critical 
amplitudes of the susceptibility  $\chi^{(-)}(u;S)$ 
 are related to ours  by
  the factor  $S^2$ (-ln$ u_c(S))^{\gamma}/u_c(S)^{4S}$.
A similar remark applies to the specific heat amplitudes 
  $C^{(-)}(u;S)$ for which the conversion factor is 
$1/u_c(S)^{4S}$(ln$ u_c(S))^{2}$. 
 Finally, the magnetization amplitudes of Ref.\cite{ejg} should be
 multiplied by the factor $($-ln$ u_c(S))^{1/8}/S$ to agree with ours.
 Of course, the amplitudes of the conformal 
field theory considered in the  study of 
Ref.\cite{delf} are not comparable to our series quantities.

 As indicated in eqs.(\ref{mag}) - (\ref{conf4}), 
 for a given spin $S$,  all asymptotic forms  are 
moreover expected\cite{weg} to contain
 leading non-analytic confluent corrections characterized by
the same exponent  $\theta(S)$.
Higher order corrections are also expected to contain 
logarithmic\cite{weg} factors.
 If universality holds, all exponents  have to be $S$-independent. 

The presence and the value of  the confluent exponent has been 
discussed\cite{ahafi,case,blonij,barfish,adler}
 several times. 
 From RG calculations\cite{zinn},  
 both in the $\epsilon$-expansion and in the fixed-dimension approach, 
it was  conjectured 
 that  $\theta \simeq 4/3$ for the  universality class of 
the two-dimensional Ising model. 
 Aharony and Fisher and later 
Bl\"ote and den Nijs argued\cite{ahafi,blonij} 
that $a^{(\pm)}_{\chi}(S)=0$ 
for $S=1/2$  and indeed no such correction was  revealed
 by the later very accurate study\cite{wu,orri} of the 
critical asymptotic expansion for $\chi(K;1/2)$.
However, in the absence of more general results, 
the  reliable assessment of the subleading asymptotic critical behaviour 
 remained an  open problem when $S>1/2$.

\section{ Estimates of universal amplitude combinations}

In terms of $\chi^{(+)}(K;S)$, $\xi^{(+)}(K;S)$  
and $ \chi^{(+)}_{4}(K;S)$, a ``hyper-universal'' combination 
of critical 
 amplitudes denoted by $ g^{(+)}_r(S)$ and usually called   the
``dimensionless renormalized
 coupling  constant'',   can be defined 
by 

\begin{equation}
  g^{(+)}_r(S)
=  - \frac{3v   C^{(+)}_{4}(S)}{8 \pi (a f^{(+)}(S))^2 C^{(+)}(S)^2}
\label{ampg}\end{equation} 

Here    the normalisation factor 
 $\frac{3} {8\pi}$ is chosen in order 
to match the usual field theoretic definition\cite{zinn} 
of $g^{(+)}_r(S)$ and  $v$ denotes the volume per lattice site,
measured in units of the square of a lattice constant. 
For all lattices
one has $v=\sigma a^2$, with $a$ the lattice constant. 
For the triangular lattice, $\sigma=\sqrt{3}/2,$ for
the honeycomb lattice $\sigma=3\sqrt{3}/4,$ and for the
kagom\'e lattice $\sigma = 2/\sqrt{3}.$

 We have also studied the hyper-universal combination
 usually denoted as
\begin{equation}
R^{(+)}_{\xi}(S)=(A^{(+)}(S)/v)^{1/2}(af^{(+)}(S)) 
\label{rxi}\end{equation} 
and   the  Watson combination\cite{wats}

\begin{equation}
 R_C(S)=A^{(+)}(S) C^{(+)}(S)/B^{(-)}(S)^2
\label{rc}\end{equation} 

The other frequently considered universal combination
 
\begin{equation}
 R_4(S)= C_4^{(+)}(S)B^{(-)}(S)^2/C^{(+)}(S)^3,
\label{r4}\end{equation}  
is not independent of the previous ones, since  
$R_4(S)=-\frac {8} {3\pi} g^{(+)}_r(S) R^{(+)}_{\xi}(S)^2/ R_C(S)$. 

All of these quantities  are accurately known in the $S=1/2$ case. 
As indicated in Ref.\cite{wu,aha}, it is known that 
$A^{(+)}(1/2)=\frac {2} {\pi}$ln$[\tan(\pi/8)]^2 \approx 0.4945385895$, 
$C^{(+)}(1/2) \approx 0.962581732 $,
and  $B^{(-)}(1/2)=2^{5/16}$ln$[1+\sqrt {2}]^{1/8} \approx 1.22240995$.
 In  Refs.\cite{saso,case}, 
we find the very accurate estimates $C_4^{(+)}(1/2)= 4.379095(8)$  
 and $f^{(+)}(1/2)\approx 0.5670683  $.

Therefore we can conclude   $g^{(+)}_r(1/2) = 1.754364(2)$, 
  $R_C(1/2) \approx 0.31856939$,  
 $R^{(+)}_{\xi}(1/2) \approx  0.39878194 $
 and $R_4(1/2)=7.336744(10)$.

 We have also considered the ratio $C^{(+)}(S)/C^{(-)}(S)$.  
 In Ref.\cite{wu}, for $S=1/2$,  this ratio was computed 
with arbitrary precision to be  
 $C^{(+)}(1/2)/C^{(-)}(1/2) \approx 37.693652$. 
 
Finally, we have estimated 
the ratio $A^{(+)}(S)/A^{(-)}(S)$   
for various values of $S.$    
This ratio equals unity for $S=1/2$  by self-duality. 
 This was argued\cite{ande} in greater generality for the 
 $q$-state ($0 \le q \le 4$) Potts model 
on the square lattice, which, for $q=2$, 
reduces to the $S=1/2$ Ising model. 

In what follows, we determine the values of these universal 
amplitude combinations for $S>1/2.$ 
The preliminary part of our series analysis  
 is  aimed at estimating  the critical temperatures using the 
expansions of  
 $\chi^{(+)}(K;S)$ for $S \geq 1$. 
 We  employed a variety of methods:
 Zinn-Justin improved-ratio formula\cite{zinn81}, 
Pad\'e approximants (PA) and  inhomogeneous 
  differential approximants (DA)\cite{gutda}. 
The best results with DA's were obtained   from  approximants  
such  that the polynomial coefficient of the highest derivative is
 even. (As a consequence, the approximants always contain
an additional anti-ferromagnetic
 singularity at $-K_c(S)$, beside the
one at $K_c(S)$). 
Similarly to the LT analysis, but to a much smaller extent, 
the accuracy of our results tends to 
 deteriorate with increasing $S$. In spite of this,  
our final HT estimates
 of the critical points,  reported in Table \ref{tab3}, 
 show  significant improvement in apparent accuracy  and 
sizable discrepancies from
 the previous LT determinations\cite{ejg}.

\begin{table}[b]
\squeezetable
\caption{ Estimates of the critical inverse temperatures 
 for the spin-$S$ Ising   models
 on the sq   lattice. Of course, the estimate $K_c^{(+)}(S)$, 
 obtained from the HT series, must equal $K_c^{(-)}(S)$ 
 obtained from the LT series, and their common value is known
exactly only for $S=1/2$. 
For comparison,  we have also reported  other  results
beside those  obtained  
from our HT and those obtained in Ref.\cite{ejg} 
 from the analysis of  LT  series. 
No error estimates are provided in Ref.\cite{burk} for the 
estimates of $K_c^{(+)}(S)$ obtained from the 
ten term series\cite{camp} of Camp and
 Van Dyke as well as for the estimates $K_c(S)$ obtained 
by a renormalization group method.} 
\label{tab3}
\begin{ruledtabular}
\begin{tabular}{lccccccc}
 &S=1/2 & S=1 & S=3/2 &  S=2&  S=5/2&S=3&  S=$\infty$ \\
\tableline
$K_c^{(+)}(S)$ &0.44068679\ldots&0.590473(5)
&0.684255(6)&0.748562(8)&0.79541(1)&0.83106(2)& 1.09315(2)\\ 
$K_c^{(-)}(S)$\cite{ejg} &0.44068679\ldots 
&0.5904727(9)&0.684338(46)&0.7487(14)&0.8025(35)&0.839(10)& \\
$K_c^{(+)}(S)$\cite{burk}&0.441&0.592&0.687&0.752&0.800&0.836\\
$K_c(S)$\cite{burk}&0.458&0.610&0.704&0.770&0.818&0.855&\\
$K_c^{(+)}(S)$\cite{bln} && 0.5904727(10)&&&&&\\
$K_c(S)$\cite{lipo} && 0.590471&&&&&\\
$K_c(S)$\cite{yuri} && 0.590076...&&&&&\\

\end{tabular}
\end{ruledtabular}
\end{table}

For general values of $S$, less accurate estimates of $K_c(S)$ 
 have been obtained in Ref.\cite{burk}  from the ten term 
susceptibility series of Ref.\cite{camp} 
and from a renormalization group
method. More recently other estimates\cite{monr}
 were obtained by a generalized cluster method.
To our knowledge other accurate  determinations of the
 critical points are available only for
$S = 1$. 
They have been obtained either by analysing\cite{adler} the 21 term 
HT series of Ref.\cite{nr90} for the susceptibility  or 
by various transfer-matrix methods\cite{bln,yuri,lipo}. 
Some of these results have also been cited in Table \ref{tab3}.  

We have then turned to the critical exponents $\gamma(S)$, $\nu(S)$ 
and $\gamma_4(S)$ and have evaluated them  
 from the log-derivatives of the appropriate  HT 
 expansions by first order DA's 
biased with our HT estimates of the critical temperatures. 
 This computation shows that the relative variation of the exponents 
 is smaller than $\approx 10^{-3}$, in the worst case,
 for $S$ varying between $1/2$ and $\infty$. 
We report these results in Table \ref{tab31} without  further details  
 and simply conclude 
that universality and hyperscaling appear to be
 well supported for the leading critical exponents.

\begin{table}
\squeezetable
\caption{ Estimates of  critical exponents obtained from our HT series  
 for the spin-$S$ Ising   models on the sq   lattice. Of course the
 values of $\gamma$, $\nu$ and  $\gamma_4$, 
for $S=1/2$ are exactly known.}
\label{tab31}
\begin{ruledtabular}
\begin{tabular}{lccccccc}
 &  S=1/2 & S=1 & S=3/2 &  S=2&  S=5/2&S=3&  S=$\infty$ \\
\tableline
$\gamma(S)$ &1.75 &1.7502(4)
&1.7500(4)&1.7496(5) &1.7500(5)&1.7501(5)&1.7494(8) \\ 
$\nu(S)$  &1.0  &0.9999(6)&0.9996(8) 
&0.9994(8)&0.9994(8)&0.9994(8)&0.9994(8) \\
$\gamma_4(S)$&5.5&5.498(4)&5.497(5)&5.497(5)&5.497(5)&5.497(5)&5.497(5)\\

\end{tabular}
\end{ruledtabular}
\end{table}

It is perhaps also worth noticing that {\it assuming}
 the universality of $\gamma$
 we can bias and therefore 
refine the determination of $K_c(S)$. This procedure does not change 
 the central values of the critical points with respect 
to the unbiased one, but reduces the error bars.

On the other hand, the estimate of the 
exponent $\theta(S)$ of the leading singular confluent 
 corrections to scaling in the various observables remains 
quite elusive. Performing either a Baker-Hunter\cite{hb}  or 
a Zinn-Justin\cite{zinn81} analysis, we can conclude that, 
at the level of accuracy made possible
 by the present extension of the HT series,   
the amplitudes of these corrections 
are very small, (or perhaps vanishing) for all values of $S$.
 We should mention that a similar conclusion was suggested for $S=1$ 
in Ref.\cite{blonij}, while the opposite conclusion was advocated
in Ref.\cite{adler}. 

Once we have estimated the critical temperatures and verified the 
universality of
the leading exponents, we can proceed with the analysis  
simply {\it assuming}  that, for all values of $S$,  
these exponents take exactly the values expected for the 
universality class 
of the two-dimensional spin-1/2 Ising model and {\it using} 
 them along with our estimated  
critical temperatures to bias the evaluation of the HT and LT critical
amplitudes defined by eqs.(\ref{mag}) - (\ref{conf4}). 

Our estimates of the  critical amplitudes 
are reported in Table \ref{tab4}.
\begin{table}[b]
\squeezetable
\caption{Our series estimates of the critical amplitudes on the HT
and the LT side of the critical point 
 for various spin-$S$ Ising   models
 on the square lattice. Along with our estimates of the LT amplitudes
 we have reported for comparison 
also the results of Ref.\cite{ejg} 
multiplied by the proper conversion factors.
The values of $C^{(+)}(1/2)$, $A^{(+)}(1/2)$,  
$C^{(-)}(1/2)$, $B^{(-)}(1/2)$ and $A^{(-)}(1/2)$ are 
exactly known, those of 
$f^{(+)}(1/2)$ and $C^{(+)}_4(1/2)$  are very accurately known. }
\label{tab4}
\begin{ruledtabular}
\begin{tabular}{lccccccc}
 &  S=1/2(ex.) & S=1 & S=3/2 &  S=2&  S=5/2&S=3&  S=$\infty$ \\
\tableline
$C^{(+)}(S)$&$0.9625817 \dots $ &0.5514(2)
&0.4307(2) &0.3755(2) &0.3441(2) &0.3254(3) &0.2351(2)\\ 
$A^{(+)}(S)$ &$0.494538589 \ldots$ &0.736(10)
&0.854(4)&0.917(4) &0.956(4)&0.983(4)&1.054(6)\\ 
$f^{(+)}(S)$ &0.567068$\ldots$ &0.4640(2)
&0.4309(2)&0.4159(2) &0.4082(2)&0.4033(2)&0.3900(2) \\ 
$C^{(+)}_4(S)$& 4.379095(8)&0.9630(4)&0.5073(3) &0.3591(3)&0.2902(3) 
&0.2533(3)& 0.1239(3) \\
\tableline
$C^{(-)}(S)$  &$0.02553697 \ldots $   &0.01462(3) & 0.0114(3)
&0.0102(6)&0.0090(8)&0.0055(30)&  \\
$C^{(-)}(S)$\cite{ejg}  &&0.01462(2)& 
0.0109(29)& 0.0094(10)  &0.0096(33)& &  \\
\tableline
$B^{(-)}(S)$   & $1.22240995 \ldots$ &1.131(4) &1.076(5) &1.041(5)&
1.016(5)&1.001(5)&  \\
$B^{(-)}(S)$\cite{ejg}   &   &1.1313(2)&1.077(9) &1.042(16)&
1.030(19)&1.016(26)&  \\
\tableline
$A^{(-)}(S)$  &$0.494538589 \ldots$& 0.738(6)&0.855(10) 
& 0.915(10)&1.1(2)&1.1(2)&  \\
$A^{(-)}(S)$\cite{ejg}  &&0.73(2) & 0.77(6)
&0.86(8)&0.87(9)&&  \\
\end{tabular}
\end{ruledtabular}
\end{table}

 We have employed quasi-diagonal non-defective PA's  or  DA's for 
extrapolating to  $K_c(S)$ the effective amplitudes of the 
susceptibility  and of the derivative of the  specific heat, 
from the HT and the LT side of
 the critical points. We have similarly studied 
the effective amplitudes
 of the correlation-length (available only in the HT region) 
 and of the magnetization. 
For proper comparison, in the same Table we
have also cited the  LT estimates of the  critical amplitudes for 
the spontaneous magnetization,  
the specific heat and the susceptibility
 previously obtained in Ref.\cite{ejg}. These quantities have been 
multiplied by the above indicated
 conversion factors to agree with our normalisation 
conventions. The uncertainties we have reported,
which  allow for the observed spreads in the approximant values, 
provide a subjective assessment of residual 
trends in the sequence of estimates 
and for the (unbiased) uncertainties of the critical points. 
The HT amplitudes can be determined with
a relative  accuracy ranging 
from  $\approx 10^{-3}$ in the case of the 
susceptibility, to  $\approx 10^{-2}$ 
in the case of the specific heat.
 The LT amplitudes are subject to larger relative 
 uncertainties, increasing with 
$S$, and reaching up to $\approx 50\%$ for $S>2$. In some cases, in
 order to improve 
 the  accuracy of the estimates of the LT amplitudes for $S \leq 2$,
  we have based our extrapolations
 only on the   data for $|t(S)| \gtrsim 0.02-0.04$. This 
 unconventional but reasonable procedure reduces 
 the sensitivity of the approximants to 
 the unphysical nearby  singularities. Unfortunately, 
 even this  prescription
 fails to work satisfactorily for $S>2$. 

Using only the HT series, we can  evaluate  $g^{(+)}_r(S)$, 
either directly, in terms 
of the amplitudes reported in Table \ref{tab4}, 
or by  extrapolating 
 to the critical points via DA's
 the HT expansion of the inverse effective coupling 
$1/g^{(+)}_r(K;S)=-\frac {8 \pi} {3}  
\xi(K;S)^2 \chi(K;S)^2/\chi_{4}(K;S)$. A third approach
 consists in studying the 
residua at $x=1$ of the series 
with coefficients $a_n(S)= c_n(S)/d_n(S)$, where 
$c_n(S)$ are the HT coefficients of $\xi^2(K;S)$ and $d_n(S)$ are 
the coefficients of the quantity $\chi_4(K;S)/\chi^2(K;S)$.
In Table \ref{tab4} we have reported 
the results of the latter procedure
 since it yields estimates with smaller spreads.

Several other estimates\cite{bb,lai,bc96,kim,case,balo} 
obtained by a 
variety of methods are also available in the literature.

Using also the LT series, we have 
 evaluated, directly in terms of the single amplitudes,    
 the other mentioned universal combinations,  
 for a range of values of $S$.
We have reported in Table \ref{tab8}, 
our series estimates of all these 
quantities for $S > 1/2$. In conclusion, 
whenever only HT amplitudes are involved,
 our estimates, within a precision up to 
$0.1 \%$, are independent of $S$, 
in full agreement with universality. On the other hand, 
our reanalysis of the
 LT series has been only partially successful: whenever LT 
amplitudes also
 enter into the combinations, universality appears to be fairly 
well respected 
 for $S<5/2$, but the uncertainties grow notably 
 larger for larger values of the spin.

\begin{table}
\squeezetable 
\caption{Universal combinations of critical amplitudes for various 
spin-$S$ Ising models on the square lattice. 
The  exactly (or very accurately) known values  for $S=1/2$
are reported in the first column.
For $ S > 1/2 $, the series estimates of this note 
 are reported in the 
successive columns. In the last line we 
have reported the estimates of 
$R_C(S)$  obtained by combining our present
estimates of $A^{(+)}(S)$ and $C^{(+)}(S)$ 
with the estimates of 
$B^{(-)}(S)$ given in Ref.\cite{ejg}. }
\label{tab8}
\begin{ruledtabular}
\begin{tabular}{lccccccc}
&   S=1/2(ex.)&  S=1 & S=3/2 &  S=2&  S=5/2&S=3 &$S=\infty$ \\
\tableline
$C^{(+)}(S)/C^{(-)}(S)  $ & $37.693652 \dots$  &37.71(9)
&38(1). &37(2). &38(3).&59(32).&\\ 
$A^{(+)}(S)/A^{(-)}(S)$  &1.0&0.997(21) 
&0.999(16)&1.0(1)&0.87(16)&0.89(17)&  \\
$g^{(+)}_r(S)$ &1.754364(2) & 1.753(2)& 1.753(2)
&1.752(3)&1.752(3)&1.752(3)&1.752(3) \\
$R^{(+)}_{\xi}(S)$  &$ 0.39878194\ldots$ & 0.398(3)&  0.398(1)
&0.398(1)&0.399(1) &0.400(2)&0.400(2) \\
$ R_C(S)$ & $0.31856939 \ldots$  & 0.317(5)&0.318(2)& 
0.318(2)&0.319(3)& 0.319(5)&  \\
$ R_C(S)$\cite{ejg}  && 0.317(4) &0.317(7)
& 0.317(11)&0.31(1) & 0.31(1)&\\
\end{tabular}
\end{ruledtabular}
\end{table}

In the next section we describe the theory of lattice-lattice scaling,
and show how it can be used to extend our estimates of the 
critical amplitudes from the square
lattice to other two-dimensional lattices.

\section  {Lattice-lattice scaling}

The theory of lattice-lattice scaling was developed by Betts,
Guttmann and Joyce\cite{bgj} in the early '70s. It 
explains how amplitudes change within a given universality
class, as one moves from one lattice to another. It can also be 
viewed as a generalisation of the law of
corresponding states. In this section we give a terse development
of the theory, and apply it to the problem at hand. 

In order to review the general ideas let us
first consider the Weiss theory or mean field theory of a magnetic
system. The equation of state is well known to be
 $$h=\frac{1}{3}m^3(1+3t/m^2).$$
Here $t=T/T_c-1,$ $h=\mu H/kT,$ and $m=M(T)/M(0)$ are the reduced
temperature, magnetic field and magnetization, respectively.

Then the {\em law of corresponding states} says that the equation
of state is the same for all lattices. That is,
$$m_{\rm X}(t,h)=m_{\rm Y}(t,h),$$ where ${\rm X}$ and ${\rm Y}$ 
denote two lattices.
That is to say,  the lattice dependence is entirely
contained in the critical temperature $T_c.$

A more complex model is the three-dimensional spherical model,
for which the critical equation of state is:
$$h = D_{\rm X}m^5(1 + t/m^2)^2.$$ Here both $T_c$ and the amplitude $D$
are lattice dependent.
Thus $$m_{\rm X}(t,h_{\rm X}) = m_{\rm Y}(t,h_{\rm Y}).$$ 
We see that we must scale the 
field variable, so that $\frac{h_{\rm X}}{D_{\rm X}} =
 \frac{h_{\rm Y}}{D_{\rm Y}},$ 
but that there is no need to scale
the reduced temperature.

Let us now consider the case of 
 the (zero-field) spin $S = 1/2$ 
Ising model on the triangular (${\rm T}$) and hexagonal (${\rm H}$) 
lattices. 
The star-triangle relation\cite{ons,fis}
 allows us to relate the free-energy, susceptibility
and spontaneous magnetization between the lattices:
\begin{eqnarray*}
2f_{\rm H}(K_{\rm H}) = f_{\rm T}(K_{\rm T}),&& \\ \nonumber
M_{\rm T,0}(t_{\rm T}) = M_{{\rm H},0}(t_{\rm H}),&& \\ \nonumber
2\chi_{\rm T}(v_{\rm T})=
\chi_{\rm H}(v_{\rm H}) + \chi_{\rm H}(-v_{\rm H}),
\end{eqnarray*}
where $f = 1/N \ln{Z},$ $K_{\rm {\rm X}} = J_{\rm X}/kT,$ 
 and $v_{\rm X} = \tanh(J_{\rm X}/kT).$
Here we see that the 
%scaled 
 reduced temperature needs to be re-scaled for 
the free-energy to be universal. This is not restricted to the 
triangular-honeycomb pair, but in that case it is easy to be explicit. 

All these examples can be encapsulated in the following expression
for the singular part of the free-energy:
$$n_{\rm X}f_{\rm X}(t_{\rm X},h_{\rm X}) = 
 n_{\rm Y}f_{\rm Y}(t_{\rm Y},h_{\rm Y}) = f(t,h),$$ where
the reduced temperature and field are scaled by 
$$n_{\rm X}h_{\rm X} = n_{\rm Y}h_{\rm Y} = h$$ and
$$g_{\rm X}t_{\rm X} = g_{\rm Y}t_{\rm Y} = t.$$  
The singular part of the free-energy, $f(t,h),$
is then a universal  (lattice independent) function for a given model.

Equivalently, by differentiation we obtain
$$m_{\rm  X}(t_{\rm X},h_{\rm X}) = m_{\rm Y}(t_{\rm Y},h_{\rm Y}) 
= m(t,h),$$ and
$$\chi_{\rm X}(t_{\rm X},h_{\rm X})/n_{\rm X} 
= \chi_{\rm Y}(t_{\rm Y},h_{\rm Y})/n_{\rm Y} = \chi(t,h),$$ where
$m(t,h)$ and $\chi(t,h)$ are universal functions for a given model.

Writing $m_{\rm X} = B^{(-)}_{\rm X}(-t_{\rm X})^\beta,$ it follows that
$$\frac{g_{\rm X}}{g_{\rm Y}} = 
\left(\frac {B^{(-)}_{\rm X}}{B^{(-)}_{\rm Y}}\right)^{1/\beta}.$$
Using this result, and the exact scaling parameters $g_{\rm X}$ 
and $g_{\rm Y}$ given below, it is a trivial matter to calculate 
the magnetization
amplitudes for the other lattices we consider (triangular, hexagonal
and kagom\'e (${\rm K}$)), taking as 
input the square lattice amplitudes given
in Table V. These amplitude estimates are given 
in Tables  VII, VIII and IX.

Writing $\chi_{\rm X} = C_{\rm X}^{(\pm)}t_{\rm X}^{-\gamma}$ 
it follows that
$$C_{\rm X}^{(+)}/C_{\rm Y}^{(+)} = 
\frac{n_{\rm X}}{n_{\rm Y}}
\left(\frac{g_{\rm X}}{g_{\rm Y}}\right)^{-\gamma}
= C_{\rm X}^{(-)}/C_{\rm Y}^{(-)}.$$
Similarly,
it is a trivial matter to calculate the susceptibility
amplitudes for the other lattices we consider, 
taking as input the $S=1/2$ square lattice amplitudes given
in Table V. 

Further differentiation gives the corresponding relationship for
higher field derivatives, and we readily obtain
$$C_{2l,{\rm X}}^{(+)}/C_{2l,{\rm Y}}^{(+)} 
= \left(\frac{n_{\rm X}}{n_{\rm Y}}\right)^{2l-1}
\left(\frac{g_{\rm X}}{g_{\rm Y}}\right)^{-\gamma_{2l}},$$
for the high-temperature field derivatives, 
(where only the even-order
derivatives are non-zero). 
The corresponding result for low-temperature
field derivatives is
$$C_{l,{\rm X}}^{(-)}/C_{l,{\rm Y}}^{(-)} 
= \left(\frac{n_{\rm X}}{n_{\rm Y}}\right)^{l-1}
\left(\frac{g_{\rm X}}{g_{\rm Y}}\right)^{-\gamma'_{l}}.$$

Taking temperature derivatives, 
one readily establishes that the specific
heat amplitudes satisfy
$$A^{(+)}_{\rm X}/A^{(+)}_{\rm Y} 
= \frac{n_{\rm Y}}{n_{\rm X}}\left(\frac{g_{\rm X}}{g_{\rm Y}}
\right)^{2-\alpha} = A^{(-)}_{\rm X}/A^{(-)}_{\rm Y}. $$ 
As $\alpha = 0$ for the two-dimensional Ising model, 
this simplifies to
$$A^{(+)}_{\rm X}/A^{(+)}_{\rm Y} 
= \frac{n_{\rm Y}} {n_{\rm X}}
\left(\frac{g_{\rm X}} {g_{\rm Y}}\right)^2 
= A^{(-)}_{\rm X}/A^{(-)}_{\rm Y}. $$ 
We have similarly calculated the specific-heat
amplitudes for the other lattices we consider, 
taking as input the square lattice amplitudes given
in Table VI. These are also given in Tables VII, VIII, and IX.
\begin{table}
\squeezetable
\caption{Estimates of the critical amplitudes on the HT
and the LT side of the critical point 
 for various spin-$S$ Ising   models
 on the triangular lattice, as obtained by lattice-lattice scaling,
using the square-lattice series estimates as a basis.
The values of $C^{(+)}(1/2)$, $A^{(+)}(1/2)$,  
$C^{(-)}(1/2)$, $B^{(-)}(1/2)$ and $A^{(-)}(1/2)$ are exactly known. }
\label{tab5}
\begin{ruledtabular}
\begin{tabular}{lccccccc}
 &  S=1/2(ex.) & S=1 & S=3/2 &  S=2&  S=5/2&S=3&  S=$\infty$ \\
\tableline
$C^{(+)}(S)$&$0.92420696 \dots $ &0.5294(2)
&0.4135(2) &0.3605(2) &0.3304(2) &0.3124(3) &0.2257(2)\\ 
$A^{(+)}(S)$ &$0.499069377 \ldots$ &0.743(10)
&0.862(4)&0.925(4) &0.965(4)&0.992(4)&1.064(6)\\ 
$f^{(+)}(S)$ &0.525315$\ldots$ &0.4298(2)
&0.3992(2)&0.3853(2) &0.3781(2)&0.3736(2)&0.3613(2) \\ 
$C^{(+)}_4(S)$& 4.000248(8)&0.8797(4)&0.4634(3) &0.3280(3)&0.2651(3) 
&0.2314(3)& 0.1132(3) \\
\tableline
$C^{(-)}(S)$  &$0.024518902 \ldots $   &0.01404(3) & 0.0109(3)
&0.0098(6)&0.0086(8)&0.0053(30)&  \\
\tableline
$B^{(-)}(S)$   & $1.203269903 \ldots$ &1.113(4) &1.059(5) &1.025(5)&
1.000(5)&0.985(5)&  \\
\tableline
$A^{(-)}(S)$  &$0.4990693724 \ldots$& 0.745(6)&0.863(10) 
& 0.923(10)&1.1(2)&1.1(2)&  \\
\end{tabular}
\end{ruledtabular}
\end{table}

\begin{table}
\squeezetable
\caption{Estimates of the critical amplitudes on the HT
and the LT side of the critical point 
 for various spin-$S$ Ising   models
 on the hexagonal lattice, as obtained by lattice-lattice scaling,
using the square-lattice series estimates as a basis.
The values of $C^{(+)}(1/2)$, $A^{(+)}(1/2)$,  
$C^{(-)}(1/2)$, $B^{(-)}(1/2)$ and $A^{(-)}(1/2)$ are exactly known. }
\label{tab6}
\begin{ruledtabular}
\begin{tabular}{lccccccc}
 &  S=1/2(ex.) & S=1 & S=3/2 &  S=2&  S=5/2&S=3&  S=$\infty$ \\
\tableline
$C^{(+)}(S)$&$1.0464170 \dots $ &0.5994(2)
&0.4682(2) &0.4082(2) &0.3741(2) &0.3537(3) &0.2556(2)\\ 
$A^{(+)}(S)$ &$0.4781063817 \ldots$ &0.712(10)
&0.826(4)&0.887(4) &0.924(4)&0.950(4)&1.019(6)\\ 
$f^{(+)}(S)$ &0.657331$\ldots$ &0.5379(2)
&0.4995(2)&0.4821(2) &0.4732(2)&0.4675(2)&0.4521(2) \\ 
$C^{(+)}_4(S)$& 5.352965(8)&1.1772(4)&0.6201(3) &0.4390(3)&0.3547(3) 
&0.3096(3)& 0.1515(3) \\
\tableline
$C^{(-)}(S)$  &$0.027761095 \ldots $   &0.01589(3) & 0.0124(3)
&0.0111(6)&0.0098(8)&0.0060(30)&  \\
\tableline
$B^{(-)}(S)$   & $1.253177691 \ldots$ &1.159(4) &1.103(5) &1.067(5)&
1.042(5)&1.026(5)&  \\
\tableline
$A^{(-)}(S)$  &$0.4781063817 \ldots$& 0.714(6)&0.827(10) 
& 0.885(10)&1.1(2)&1.1(2)&  \\
\end{tabular}
\end{ruledtabular}
\end{table}

\begin{table}
\squeezetable
\caption{Estimates of the critical amplitudes on the HT
and the LT side of the critical point 
for various spin-$S$ Ising models
on the kagom\'e lattice, as obtained by lattice-lattice scaling,
using the square-lattice series estimates as a basis.
The values of $C^{(+)}(1/2)$, $A^{(+)}(1/2)$,  
$C^{(-)}(1/2)$, $B^{(-)}(1/2)$ and $A^{(-)}(1/2)$ are exactly known.}
\label{tab7}
\begin{ruledtabular}
\begin{tabular}{lccccccc}
 &  S=1/2(ex.) & S=1 & S=3/2 &  S=2&  S=5/2&S=3&  S=$\infty$ \\
\tableline
$C^{(+)}(S)$&$1.01814223 \dots $ &0.5832(2)
&0.4556(2) &0.3972(2) &0.3640(2) &0.3442(3) &0.2487(2)\\ 
$A^{(+)}(S)$ &$0.4800615653 \ldots$ &0.714(10)
&0.829(4)&0.890(4) &0.928(4)&0.954(4)&1.023(6)\\ 
$f^{(+)}(S)$ &0.618474$\ldots$ &0.5061(2)
&0.4700(2)&0.4536(2) &0.4452(2)&0.4399(2)&0.4254(2) \\ 
$C^{(+)}_4(S)$& 5.046953(8)&1.1099(4)&0.5847(3) &0.4139(3)&0.3345(3) 
&0.2919(3)& 0.1428(3) \\
\tableline
$C^{(-)}(S)$  &$0.02701097 \ldots $   &0.01546(3) & 0.0121(3)
&0.0108(6)&0.0095(8)&0.0058(30)&  \\
\tableline
$B^{(-)}(S)$   & $1.238655888 \ldots$ &1.146(4) &1.090(5) &1.055(5)&
1.030(5)&1.014(5)&  \\
\tableline
$A^{(-)}(S)$  &$0.4800615653 \ldots$& 0.716(6)&0.830(10) 
& 0.888(10)&1.1(2)&1.1(2)&  \\
\end{tabular}
\end{ruledtabular}
\end{table}

Finally, the correlation function amplitudes were calculated
exactly from various star-triangle transformations 
by Thompson and Guttmann
\cite{TG75} for the {\em true} correlation length. It follows from
universality that these same transformations should hold also for the
second-moment correlation length amplitudes. We show below that 
this is equivalent
to the universality of $R_\xi^{(+)}(S),$ 
which is a conclusion of this
work.

In \cite{TG75} it was shown by explicit calculation that
$$f_{\rm T}^{(+)}(1/2)K_{c,{\rm T}}^{(+)}(1/2)/
f_{\rm H}^{(+)}(1/2)K_{c,{\rm H}}^{(+)}(1/2)=1/3.$$
$$f_{\rm S}^{(+)}(1/2)K_{c,{\rm S}}^{(+)}(1/2)/
f_{\rm T}^{(+)}(1/2)K_{c,{\rm T}}^{(+)}(1/2)=\sqrt{3}.$$
$$f_{\rm K}^{(+)}(1/2)K_{c,{\rm K}}^{(+)}(1/2)/
f_{\rm H}^{(+)}(1/2)K_{c,{\rm H}}^{(+)}(1/2)=2/3.$$

The universality of 
$R_\xi^{(+)}(S) =\left(\frac {A^+(S)} {v}\right)^{1/2}af^+(S),$
taken together with the above lattice-lattice scaling relation for the
specific heat amplitude $A^+(S)$ implies the lattice-lattice 
scaling relation $$f^{(+)}_{\rm X}/f^{(+)}_{\rm Y} 
= \frac{g_{\rm Y}}{g_{\rm X}}\left(\frac{\sigma_{\rm X} n_{\rm X}}
{\sigma_{\rm Y} n_{\rm Y}}\right)^{1/2}, $$ 
where $\sigma$ is the area per site, and
$\sigma_{\rm X} = 1, \sqrt{3}/2, 3\sqrt{3}/4$ and $2/\sqrt{3}$
for the square, triangular, honeycomb and kagom\'e lattices respectively.
This is equivalent to the explicit amplitude scaling 
reported in \cite{TG75},
and confirms the expectation that the {\em true} correlation 
length amplitude
and the second-moment correlation length amplitude scale similarly.

For the 2d Ising model we can calculate 
the scaling parameters $n_{\rm X}$
and $g_{\rm X}$ exactly from known spin-1/2 spontaneous magnetization
and specific heat amplitudes. The critical points are also exactly
known. These are given in the table below:

\begin{tabbing}
xxxxxxxx \= xxxxxxxxxxxxxxxxx \= xxxxxxxxxxxx \= xxxxxxxxxxxxx  \kill
${\rm X}$    \>   $K_c$ \> $n_{\rm X}$   \>  
  $g_{\rm X}K_{c,{\rm T}}/K_{c,{\rm X}}$   \\
Tr. \>  $\ln(3)/4$ \> 1 \> 1 \\
Sq. \> $-\ln(\sqrt{2}-1)/2$ \> $3\sqrt{3}/4$  \> $1/\sqrt{2}$ \\
Ho. \> $-\ln(2-\sqrt{3})/2$ \> 2 \>  $1/\sqrt{3}$ \\
Ka. \> $-\ln(2/\sqrt{3}-1)/4$ \> $3^5(\sqrt{3}-1)^{16}$ 
\> $9(\sqrt{3}-1)^8$ \\
\end{tabbing}

In addition to the above results, one can also derive scaling
relations for the amplitudes of sub-dominant singularities.
For example, consider the susceptibility of the 2d Ising model 
on lattice X. Writing
$$\chi_{\rm X} \sim C_0^{\rm X} t^{-\gamma}+C_1^{\rm X} t^{1-\gamma},$$
from lattice-lattice scaling we can 
derive the following amplitude relations:
$$\frac {C_0^{\rm X}}{C_0^{\rm T}} = 
\frac {n_{\rm X}}{n_{\rm Y}}
\left(\frac{g_{\rm Y}}{g_{\rm X}}\right)^\gamma,$$
$$\frac {C_1^{\rm X}}{C_1^{\rm T}} = \frac{n_{\rm X}}{n_{\rm Y}}
\left(\frac{g_{\rm Y}}{g_{\rm X}}\right)^{\gamma-1}.$$
The second expression is false for the kagom\'e lattice. It is corrected
 by the theory of extended lattice-lattice scaling developed by 
Gaunt and Guttmann\cite{gg}
 in 1978. In that theory, a third scaling parameter needs 
to be introduced.

It is widely accepted, 
and in complete agreement with the results of the first 
part of this paper,
 that the spin-$S$ Ising model is in the same
universality class as the spin-$1/2$ Ising model. This is the only
assumption we require in order to apply the theory of lattice-lattice
scaling to the square lattice data given in the previous section.
The only subtlety is whether ``universality" really 
extends to lattice-lattice
universality. While this assumption seems natural, we did attempt to
verify it by estimating amplitudes for other lattices from the 
rather short
data available in \cite{camp}. The longest effective series is the
triangular lattice series. We found the high-temperature susceptibility
amplitude, as estimated by Pad\'e approximants, to be $C^{(+)}(1)=0.529$
from this series, in complete agreement with the more precise value
$C^{(+)}(1)=0.5294(2)$
found by lattice-lattice scaling, and reported in Table VII.

Note too that there is no loss of accuracy, as the conversions 
from lattice to lattice are exact. For example, based on the recent
estimate of the leading susceptibility amplitude of the spin $1/2$
square lattice Ising model given in\cite{orri}, application of
lattice-lattice scaling gives the corresponding amplitude on the
triangular lattice to the same precision, viz:
$$C_+^T=0.9242069582451643296971575778559317176696261520028389.$$
Similarly accurate results for other lattices can be readily written
down, as can equally accurate sub-dominant amplitudes.

\acknowledgments

This work was completed before the untimely death of Marco Comi, 
our lifelong  friend and coauthor. 
 To his dear memory we dedicate this paper.  

This work has been partially supported by the Ministry of Education, 
University and Research (PB, MC), 
and the Australian Research Council (AJG).  

%\begin{references}

\end{document}